\documentclass[aps,prd,twocolumn,superscriptaddress,nofootinbib,floatfix]{revtex4-2}

\usepackage{amsmath,amssymb,bm}
\usepackage{graphicx}
\usepackage{array,dcolumn,multirow}
\usepackage{xcolor}
\usepackage{mathrsfs}
\usepackage{comment}
\usepackage{subfigure}
\usepackage{orcidlink}
\definecolor{myred}{RGB}{179, 27, 27}
\usepackage{hyperref}
\hypersetup{
    colorlinks=true,
    linkcolor=myred,
    citecolor=myred,
    urlcolor=myred
 }

\begin{document}
\title{Geodesic stability in Schwarzschild black hole with string cloud background}
\author{Shobhit Giri \orcidlink{0000-0002-7444-0389}}
\email{shobhit6794@gmail.com}
\affiliation{Department of Physics, Swami Shraddhanand College (University of Delhi), Alipur-110036, Delhi, India.}

\author{Hemwati Nandan \orcidlink{0000-0002-1183-4727}}
\email{hnandan@associates.iucaa.in}
\affiliation{Department of Physics, Hemwati Nandan Bahuguna Garhwal University,
Srinagar--246174, Uttarakhand, India.}
\affiliation{Centre for Space Research, North-West University, Potchefstroom 2520, South Africa.}

\author{ Amare Abebe \orcidlink{0000-0001-5475-2919}}
\email{amare.abebe@nithecs.ac.zaz }
\affiliation{Centre for Space Research, North-West University, Potchefstroom 2520, South Africa.}
\affiliation{National Institute for Theoretical and Computational Sciences (NITheCS), South Africa.}

\author{ Saswati Roy \orcidlink{0000-0002-7028-2627}}
\email{sr.phy2011@yahoo.com}
\affiliation{Department of Physics, National Institute of Technology, Agartala, Tripura--799046, India.}

\author{Asifa Ashraf \orcidlink{0000-0001-5522-5491 }}
\email{asifamustafa3828@gmail.com}
\affiliation{School of Mathematical Sciences Zhejiang Normal University, Jinhua 321004, China.}
\affiliation{Center for Theoretical Physics, Khazar University, 41 Mehseti Street, Baku, AZ1096, Azerbaijan.}

\begin{abstract}
In the present work, the stability of timelike and null geodesics is studied in the vicinity of a Schwarzschild black hole with a string cloud (SCSBH) background. For this purpose, we employ the Lyapunov exponent criterion to examine the effect of the string cloud parameter on geodesic stability. We begin by analysing the dynamics of test particles around the Schwarzschild string cloud spacetime.  The effective potential for timelike and null geodesics is examined for different values of the string cloud parameter. Each fixed point is characterised according to its Lyapunov stability. Furthermore, the circular geodesics at these fixed points are classified as either stable centres or unstable saddle points, which are illustrated through the corresponding phase portraits.  Additionally, we compare the results obtained for the SCSBH with those of the Schwarzschild black hole (SBH) in the limit of vanishing string cloud parameter.
\end{abstract}

\maketitle

\section{Introduction}
\label{introduction}
In the present theoretical development, the fundamental building blocks of nature are one–dimensional extended objects (so-called strings) instead of point objects \cite{Becker:2006dvp, HenryTye:2006uv}. Indeed, the universe can be considered as a collection of the most popular candidates for such fundamental objects (extended non-point objects and one-dimensional strings). Such strings were stretched during the epoch known as inflation, causing a fast accelerated expansion of the universe, which gave rise to a large-scale object called a cosmic string \cite{Hindmarsh:1994re,Mitchell:1987th}.\\
The occurrence of phase transition led to symmetry breaking and the formation of topological defects during the development of our universe \cite{Kibble:1976sj}. Symmetry breaking occurred below a particular temperature of the universe, called the critical temperature. Strings are topological defects that arise due to symmetry breaking and may have linelike, cylindrical, and axially symmetric structures. However, one-dimensional cosmic strings are the topic of interest in most studies among all possible types of defects \cite{Vilenkin:1981iu}.
According to the string cloud model proposed by Letelier \cite{Letelier:1979ej}, an aggregation of one-dimensional objects in a defined geometric pattern is known as a string cloud. Therefore, the study of the gravitational effects of matter in the form of string clouds deserves considerable attention.
Since then, string clouds in the context of General Relativity and other alternative theories of gravity have been extensively studied \cite{Richarte:2007bx}. Exact black hole solutions with string cloud backgrounds have been investigated in general relativity \cite{Letelier:1979ej, Mazharimousavi:2009tp, Larsen:1997ge, Batool:2017bjw, Sharif:2015oua}, in Einstein-Gauss-Bonnet models \cite{Herscovich:2010vr}, and in Lovelock gravity \cite{Ghosh:2014dqa}. In the context of general relativity, the SBH for the string cloud model was obtained in the pioneering work of Letelier \cite{Letelier:1979ej}. Additionally, null geodesics and strong gravitational lensing around a BH in a string cloud background have been studied \cite{Sharif:2015oua}.  The accretion process onto the BH with a string cloud background was studied to observe the string cloud effect on the mass accretion rate \cite{Ganguly:2014cqa}. The study of spherically symmetric thin-shell wormholes in a string cloud background was carried out by Richarte and Simeone \cite{Richarte:2007bx}. A BH solution in Einstein-Gauss-Bonnet theory for the string cloud model was obtained by Herscovich and Richarte \cite{Herscovich:2010vr}. Lee et al. \cite{Lee:2014dha} found an exact static, spherically symmetric BH solution in third-order Lovelock gravity with a string cloud background. A spherically symmetric string cloud configuration in a d-dimensional spacetime within the framework of f(R) theories of gravity was derived in \cite{MoraisGraca:2017nlv}.\\
The motion of a test particle or geodesic motion in the vicinity of a BH provides insight into the geometric structure of spacetime near the BH. Various astrophysical phenomena arise due to the strong gravitational field, which makes it difficult for particles to maintain stable orbits. Therefore, it is interesting to analyse the stability of circular geodesics around a BH. In particular, we adopt the Lyapunov exponent approach, which links non-linear Einstein’s GR with non-linear dynamics \cite{AMLyapunov,PhysRevLett.55.1082,Cardoso:2008bp}. The Lyapunov exponent (or Lyapunov characteristic exponent) measures the average rate of separation between two nearby trajectories (positive for divergence and negative for convergence). 
The present study aims to investigate the stability of geodesics around the SCSBH spacetime using the Lyapunov exponent criterion (as discussed in \cite{Abolghasem2013STABILITYOC} and references therein).
Boehmer et al. \cite{boehmer2012jacobi} provided a comprehensive discussion of two important approaches for studying the stability of dynamical systems: the Lyapunov stability analysis \cite{Abolghasem2013STABILITYOC} and the Jacobi stability analysis \cite{abolghasem2012jacobi}, also known as the Kosambi–Cartan–Chern (KCC) theory \cite{kosambi2016parallelism,cartan2016observations}. These methods are widely employed in gravitational and cosmological studies, where stability plays a fundamental role in understanding the evolution of physical systems. The stability of a dynamical system is typically assessed by examining the behavior of its equilibrium (steady-state) solutions. In the Lyapunov framework, linear stability is investigated by linearizing the nonlinear system around an equilibrium point and analyzing the associated Jacobian matrix. \\
In contrast to Jacobi stability analysis, the Lyapunov approach offers a more straightforward framework for studying stability, since it involves linearizing the nonlinear dynamical system around an equilibrium (fixed) point. This feature makes the investigation of stability under small perturbations more tractable. Both Lyapunov and Jacobi stability analyses have been widely explored and applied to assess the stability of circular orbits in the neighborhood of the SBH spacetime by Hossein \cite{Abolghasem2013STABILITYOC}. Nevertheless, the Jacobi stability of dynamical systems based on the KCC theory has been extensively investigated in a variety of physical and mathematical contexts \cite{harko2008jacobi,sabau2005some,balan2009berwald,boehmer2010nonlinear}. The stability of circular orbits around a traversable wormhole with a massless conformally coupled scalar field is investigated via, Lyapunov exponent criteria \cite{Giri:2022fjy}. Furthermore, we visualize the phase-portraits to characterise each circular orbit as either a stable centre or an unstable saddle point and to examine the effect of string clouds on their stability.\\
The paper is organized as follows: In Section 2, we review the SCSBH spacetime geometry and its geodesics. In subsections A and B, we discuss the stability of timelike and null geodesics, respectively, via Lyapunov exponent and phase-portrait analysis.  The conclusion is presented in Section 3.

\section{Schwarzschild black hole with a string cloud (SCSBH) and geodesics}
\begin{figure}[]
	\centering
	{\includegraphics[height=7cm,width=8cm]{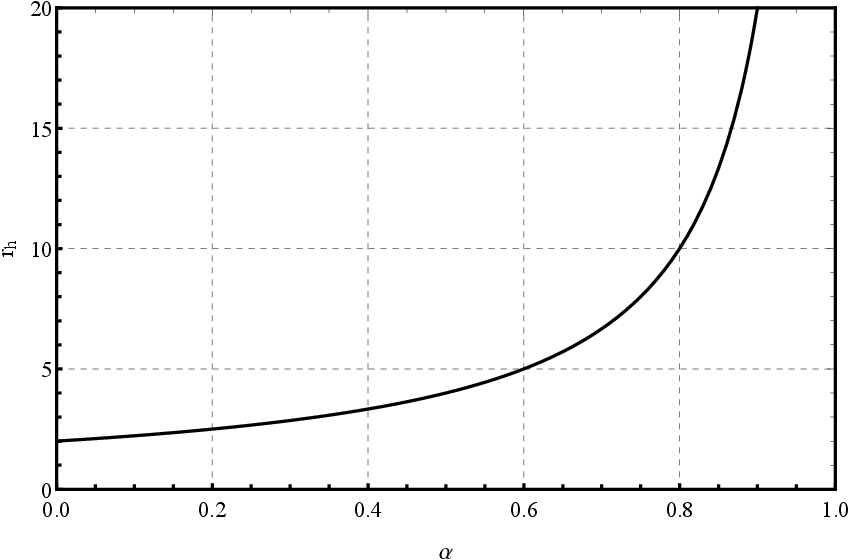}}
	\caption { Variation of event horizon radius with string cloud parameter $\alpha$ for fixed $M=1$. }\label{EHorizon}
\end{figure}

Within the framework of the string cloud model, the spacetime geometry is taken to be static and spherically symmetric, and the accretion process is assumed to be steady-state onto a nonrotating BH of mass ($M$). The string cloud theory to achieve a modified SBH with a string cloud is discussed as \cite{Letelier:1979ej,Herscovich:2010vr}. The action for such a string evolving in spacetime is known as the Nambu-Goto action, given as
\begin{multline}
S= \int_{\Sigma} \mathscr{L} d\lambda^{0} d\lambda^{1}= \int_{\Sigma}  m (-\gamma)^{1/2} d\lambda^{0} d\lambda^{1}, ~~~ \\ \gamma_{ab}= g_{\mu\nu} \frac{\partial x^{\mu}}{\partial \lambda^{a}} \frac{\partial x^{\nu}}{\partial \lambda^{b}},
\end{multline}
where $m$ denotes a positive constant associated with the string tension,
$d\lambda^{0}$, $d\lambda^{1}$
representing the timelike and spacelike parameters, respectively, and $\gamma$ represents the determinant of the induced metric $\gamma_{ab}$ on
the string world sheet $\Sigma_{ab}$.
The bivector associated with the string world sheet has a form
\begin{equation}
\Sigma^{\mu\nu}= \epsilon^{ab} \frac{\partial x^{\mu}}{\partial \lambda^{a}} \frac{\partial x^{\nu}}{\partial \lambda^{b}},
\end{equation}
where $\epsilon^{ab}$ is the two-dimensional Levi-Civita tensor and the Lagrangian density $\mathscr{L}$ for this setup takes the form of
\begin{equation}
\mathscr{L} = m \left[-\frac{1}{2}\Sigma^{\mu\nu} \Sigma_{\mu\nu}\right]^{1/2}.
\end{equation}
For a single string, the energy-momentum tensor takes the form
\begin{equation}
T^{\mu\nu} = 2 \frac{\partial \mathscr{L}}{\partial g^{\mu \nu}} =m \Sigma^{\mu\xi} \Sigma^{\nu}_{\xi}(-\gamma)^{-1/2}.
\end{equation}
The corresponding energy-momentum tensor for a cloud of strings is obtained as
\begin{equation}
T^{\mu\nu} = \rho \Sigma^{\mu\xi} \Sigma^{\nu}_{\xi}(-\gamma)^{-1/2},
\end{equation}
where $\rho$ represents the proper density of a string cloud. The quantity $\rho (-\gamma)^{-1/2}$ is referred to as the gauge-invariant density of the string cloud. In the case of a static, spherically symmetric configuration, this quantity takes the form
\begin{equation}
\rho (-\gamma)^{-1/2} = \frac{\alpha}{r^{2}}, \label{GIdensity}
\end{equation}
with $\alpha$ a positive constant, referred to as the string cloud parameter.\\
The general solution of Einstein’s field equations for a string cloud with density \eqref{GIdensity} is represented by the SCSBH metric \cite{Herscovich:2010vr}
\begin{multline}
ds^{2}= - \left[1-\frac{2M}{r}-\alpha\right] dt^{2} + \left[1-\frac{2M}{r}-\alpha\right]^{-1}dr^{2} \\+r^{2}\left[d\theta^{2}+\sin^{2}\theta d\phi^{2}\right], 
\end{multline}
where $\alpha$ is restricted to $0 < \alpha <1$ for a realistic model. 
The event horizon of the SCSBH has a radius
\begin{equation}
r_{h}= \frac{2M}{1-\alpha}, ~~ \alpha \neq 1.
\end{equation}
\noindent As  $\alpha \to 0$, the Schwarzschild radius is recovered, whereas the event horizon expands toward infinity as  $\alpha$ approaches unity. In contrast, a cloud of strings by itself ($M = 0$) does not form an event horizon; instead, it produces only a naked singularity at $r = 0$. The variation of the event horizon radius of the SCSBH spacetime with respect to the string cloud parameter ($\alpha$) is illustrated in \figurename{\ref{EHorizon}}. From the figure, it is evident that the effect of the string cloud is significantly weak for small values of $\alpha$ but the spacetime is highly distorted when the string cloud parameter approaches its limiting value.

\noindent The Lagrangian formalism is employed to study the geodesic motion of test particles confined to the equatorial plane of the aforementioned spacetime. In the equatorial plane (for $\theta=\frac{\pi}{2}$), the Lagrangian is read as
\begin{equation}
2\mathcal{L}= - \left[1-\frac{2M}{r}-\alpha\right] \dot{t}^{2} + \left[1-\frac{2M}{r}-\alpha\right]^{-1}\dot{r}^{2}+r^{2} \dot{\phi}^{2}\;.
\end{equation}
Here, an overdot represents differentiation with respect to the affine parameter $\tau$.
The spacetime’s time-translational and rotational symmetries imply conserved generalized momenta such as $p_t$ and $p_{\phi}$. Consequently, the Euler–Lagrange equations yield the first integrals of motion for $t$ and $\phi$ per unit test-particle mass as

\begin{equation}
\dot{t}= \frac{E}{\left[1-\frac{2M}{r}-\alpha\right]},\label{tdot}
\end{equation}

\begin{equation}
\dot{\phi}= \frac{L}{r^{2}}\;,\label{phidot}
\end{equation}
where quantities energy $(E)$ and angular momentum $(L)$ are conserved.
The test particle motion around spacetime is constrained by the equation
\begin{equation}
- \left[1-\frac{2M}{r}-\alpha\right] \dot{t}^{2} + \left[1-\frac{2M}{r}-\alpha\right]^{-1}\dot{r}^{2}+r^{2} \dot{\phi}^{2}= \delta . \label{constraint}
\end{equation}
The timelike and null geodesics can be obtained for $\delta=-1$ and $\delta=0$, respectively.
Substitution of Eqs.\eqref{tdot} and \eqref{phidot} into constraint Eq.\eqref{constraint} yields the radial equation governing geodesic motion in the spacetime background as
\begin{equation}
\dot{r}^{2} = E^{2} - \left(\frac{L^{2}}{r^{2}}-\delta\right) \left[1-\frac{2M}{r}-\alpha\right]\label{radialeq}.
\end{equation}
In the context of Newtonian mechanics, the motion of the test particle governed by Eq.\eqref{radialeq} may be viewed as that of a particle with energy $(\tilde{E}=E^{2}/2)$ evolving in an effective potential $(V_{eff})$. Therefore, the radial equation  above can be identified as \cite{Abolghasem2013STABILITYOC}
\begin{equation}
\frac{\dot{r}^{2}}{2}= \tilde{E}- V_{eff}(r),\label{cfe}
\end{equation}
where $V_{eff}= \frac{1}{2}\left(\frac{L^{2}}{r^{2}}-\delta\right) \left[1-\frac{2M}{r}-\alpha\right]$.
The derivative of Eq. \eqref{cfe} with respect to the affine parameter $\tau$, gives the one-dimensional radial equation $\ddot{r}= -  V_{eff}'(r)$; the prime notation $(')$ represents differentiation with respect to the radial coordinate.

To carry out the Lyapunov stability analysis of geodesic motion, the one-dimensional radial equation is reformulated as a system of first-order differential equations in the $r-p$ phase space \cite{Abolghasem2013STABILITYOC}, given by
\begin{equation}
\dot{r}= p ~, ~ \dot{p} = V_{eff}'(r).\label{rpphase}
\end{equation}
\noindent Therefore, the radial equation \eqref{cfe} is transformed into following form
\begin{equation}
\frac{{p}^{2}}{2} + V_{eff}(r)= \tilde{E}.\label{eq22}
\end{equation}
Eq. \eqref{eq22} can be linearized using Jacobi stability analysis of a vector field $F(r, p)$ corresponding to the system of Eqs. \eqref{rpphase} and the Jacobian matrix of $F$ at any point $(r, p)$ comes out as
\begin{equation}
J=  \begin{bmatrix}
0 & 1\\
-V_{eff}''(r) &  0
\end{bmatrix}.
\end{equation} 

The eigenvalue of Jacobian $J$ is called the Lyapunov exponent \cite{Abolghasem2013STABILITYOC} and is obtained as
\begin{equation} \lambda=\pm \sqrt{-V_{eff}''(r)}. \label{eigen}
\end{equation}

If the first derivative of the effective potential $V_{eff}'(r)$ vanishes at an equilibrium point or a fixed point $(r_{0}, 0)$, the fixed point  $(r_{0}, 0)$ is characterised as an unstable saddle point corresponding to $V_{eff}''(\bar{r_{0}}) < 0 $ for $\lambda$ to be real and a possible stable centre corresponding to $V_{eff}''(\bar{r_{0}}) > 0 $ for $\lambda$ to be imaginary.

\subsection{Stability of Timelike Geodesics}
The stability of timelike circular geodesics of massive test particles can be tested by taking $\delta=-1$ in Eq. \eqref{radialeq}. The effective potential for the timelike case is then given as
\begin{equation}
V_{eff}^{T}(r)=  \frac{1}{2}\left(1+\frac{L^{2}}{r^{2}}\right) \left[1-\frac{2M}{r}-\alpha\right]  .\label{effT}
\end{equation}

\begin{figure*}[htbp]
	\centering
	\subfigure[]{\includegraphics[scale=0.9]{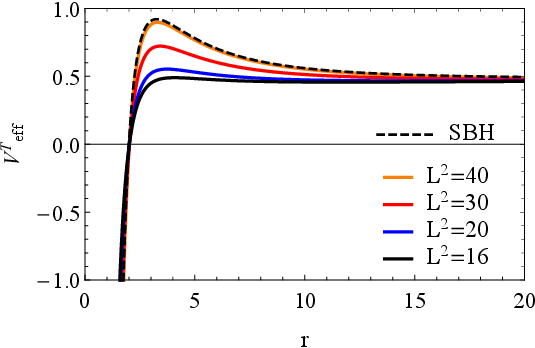}\label{ef1}}
	\subfigure[]{\includegraphics[scale=0.9]{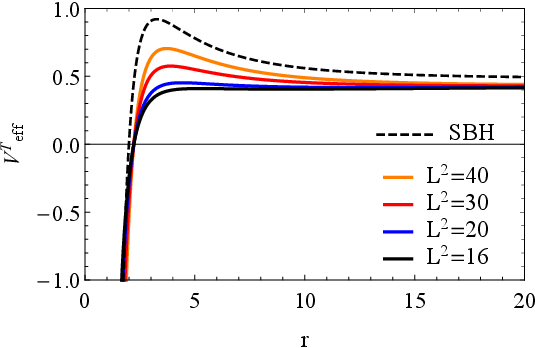}\label{ef2}}
	\caption {Variation of the effective potential $V_{eff}^{T}(r)$  with radial coordinate $r$ for timelike geodesics for string cloud parameter (a) $\alpha =0.01$, (b) $\alpha =0.1$, and for SBH considering $L^{2}=40$. Here we fixed mass $M=1$. }\label{Effectivetime}
\end{figure*}

The behaviour of the effective potential as a function of the radial coordinate is shown in \figurename{\ref{Effectivetime}} depending on the angular momentum. The effective potential increases as the angular momentum increases, whereas with the increase of the string cloud parameter, the maxima of the effective potential decrease. One can also note that as $\alpha$ reduces to zero, the effective potential of SBH becomes maximum for a particular value of angular momentum. This \figurename{\ref{Effectivetime}} also indicates that a small contribution of string cloud ($\alpha=0.01$) does not change the spacetime significantly, for which the SBH curve is very close to the SCSBH curve for $L^2=40$ but for $\alpha=0.1$, the cloud of strings modifies the spacetime geometry appreciably causes to reduce the potential barrier and approach towards the BH compared to SBH.
\\\\
\noindent For the timelike effective potential, taking $\left(V_{eff}^{T}(r)\right)'=0$ the fixed points are obtained as shown in \tablename{ \ref{table1}} and \tablename{ \ref{table11}} for the string cloud parameter $\alpha=0.01$ and $\alpha=0.1$ respectively.
\begin{table}[]
	\caption{Fixed points  ($r_{0},0$) of $V_{eff}^{T}(r)$ for timelike geodesics for $\alpha = 0.01$.}
	\centering
	\begin{tabular}{|c|c|c|c|c|}
		\hline
		Angular  & $L^{2}=40$ &  $L^{2}=30$ & $L^{2}=20$ & 	$L^{2}=16$  \\
        momentum & & & & \\[0.5ex]
		\hline
		Fixed points & 3.30636 & 3.42536 & 3.73477 & 4.0825  \\

		& 36.2936 & 26.2746  & 16.0652 & 11.7575  \\
		\hline
	\end{tabular}
	\label{table1}
\end{table}

\begin{table}[]
	\caption{Fixed points  ($r_{0},0$) of $V_{eff}^{T}(r)$ for timelike geodesics for $\alpha = 0.1$.}
	\centering
	\begin{tabular}{|c|c|c|c|c|}
		\hline
		Angular & $L^{2}=40$ &  $L^{2}=30$ & $L^{2}=20$ & 	$L^{2}=16$  \\
         momentum & & & & \\ [0.5ex]
		\hline
		Fixed points & 3.71714 & 3.8953 & 4.41742 & 5.24041  \\

		& 32.2829 & 23.1047  & 13.5826 & 9.15959  \\
		\hline
	\end{tabular}
	\label{table11}
\end{table}

\begin{table*}[tbhp]
	\caption{ Lyapunov stability of timelike geodesics around fixed points ($r_{0},0$) for different values of angular momentum and string cloud parameter.}
	\centering
	\begin{tabular}{|c|c|c|c|c|}
		\hline
		Case & ($r_{0},0$) & $\left(V_{eff}^{T}(r_{0})\right)''$ & $\lambda_{T}$  & 	Lyapunov stability  \\  [0.5ex]
		\hline
		$\alpha=0.01$, $L^{2}=16$	& 4.0825 & -0.02762 & Real & Unstable saddle \\
		
		&  11.7575 & 0.0004016 & Imaginary & Stable centre \\
		\hline
		$\alpha=0.01$, $L^{2}=20$	& 3.73477 & -0.06337 & Real &  Unstable saddle \\

		& 16.0652 & 0.000185 & Imaginary & Stable centre  \\
		\hline
		$\alpha=0.01$, $L^{2}=30$	& 3.42536 & -0.165976 & Real & Unstable saddle \\
		
		&  26.2746 & 0.0000479435 & Imaginary & Stable centre \\
		\hline
		$\alpha=0.01$, $L^{2}=40$	& 3.30636 & -0.276025 & Real & Unstable saddle \\
		
		&  36.2936 & 0.000019012 & Imaginary & Stable centre \\
		
		\hline
		$\alpha=0.1$, $L^{2}=16$	& 5.24041 & -0.005196 & Real & Unstable saddle \\
		
		&  9.15959 & 0.0005567 & Imaginary & Stable centre \\
		\hline
		$\alpha=0.1$, $L^{2}=20$	& 4.41742 & -0.024069 & Real & Unstable saddle \\

		&  13.5826 & 0.0002692 & Imaginary & Stable centre \\
			\hline
			$\alpha=0.1$, $L^{2}=30$	& 3.8953 & -0.0834365 & Real & Unstable saddle \\
			
			&  23.1047 & 0.0000262999 & Imaginary & Stable centre \\	
		\hline
		$\alpha=0.1$, $L^{2}=40$	& 3.71714 & -0.149627 & Real & Unstable saddle \\
		
		&  32.2829 & 0.0000262999 & Imaginary & Stable centre \\
		
		\hline
	\end{tabular}
	\label{table2}
\end{table*}

\begin{figure*}[htbp]
 	\centering
 	\subfigure[]{\includegraphics[scale=0.5]{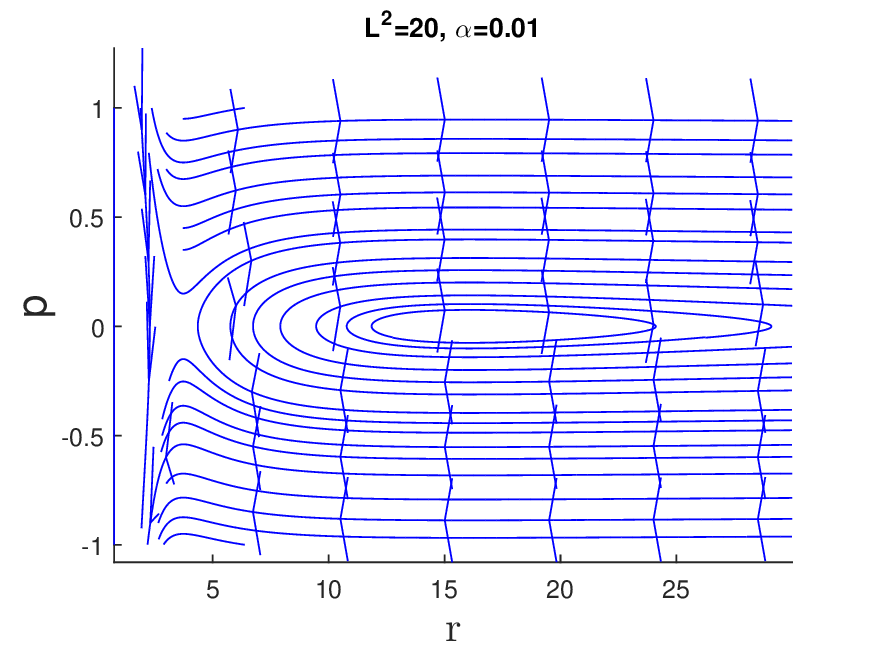}\label{ppt1}}
 	\subfigure[]{\includegraphics[scale=0.5]{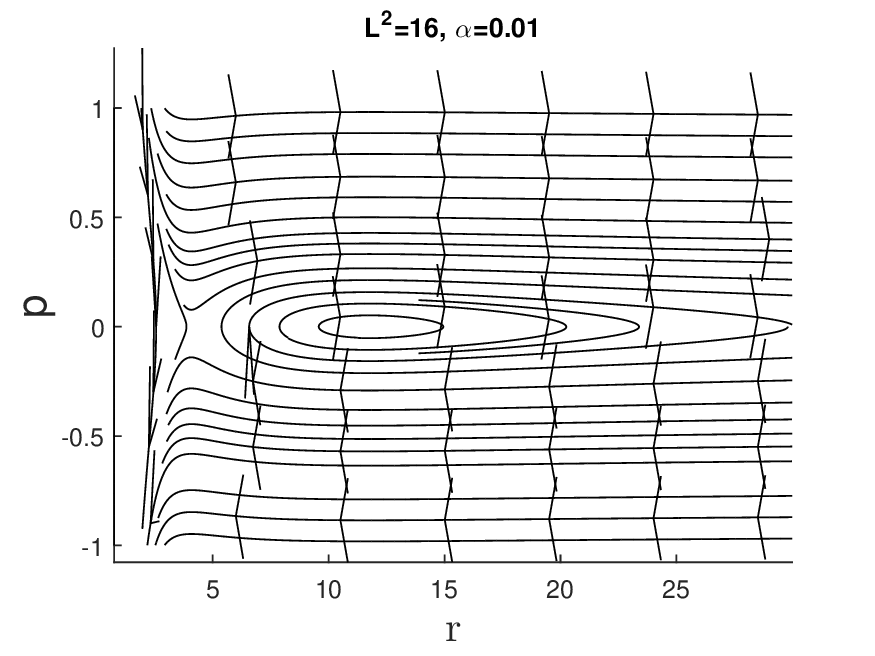}\label{ppt2}}
 	\subfigure[]{\includegraphics[scale=0.5]{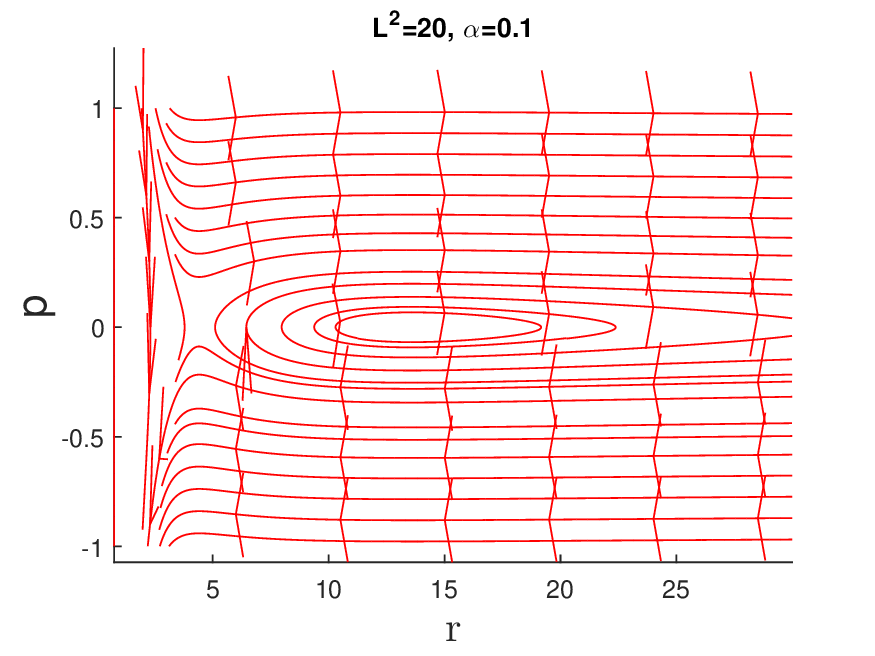}\label{ppt3}}
 	\subfigure[]{\includegraphics[scale=0.5]{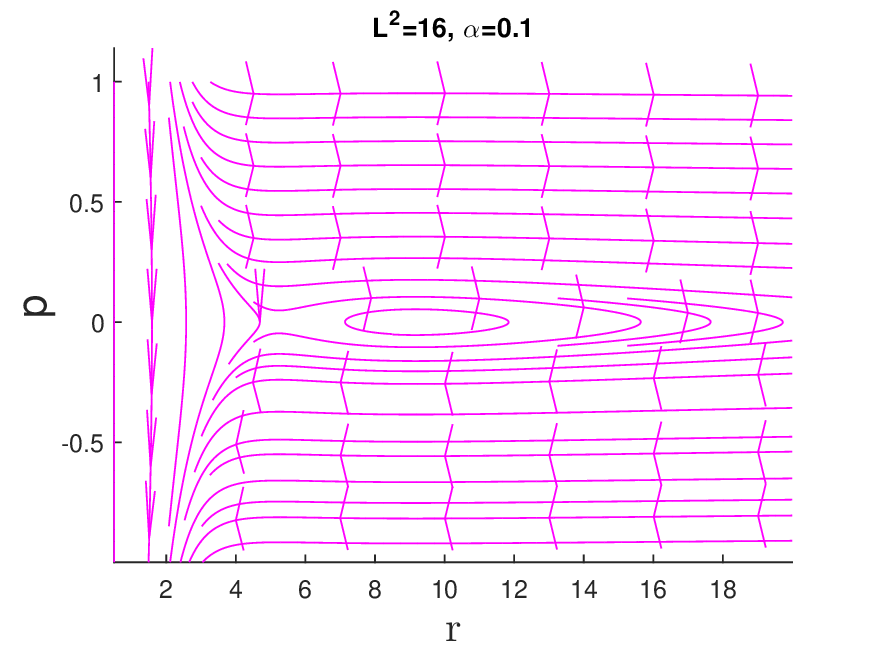}\label{ppt4}}
 	\subfigure[]{\includegraphics[scale=0.5]{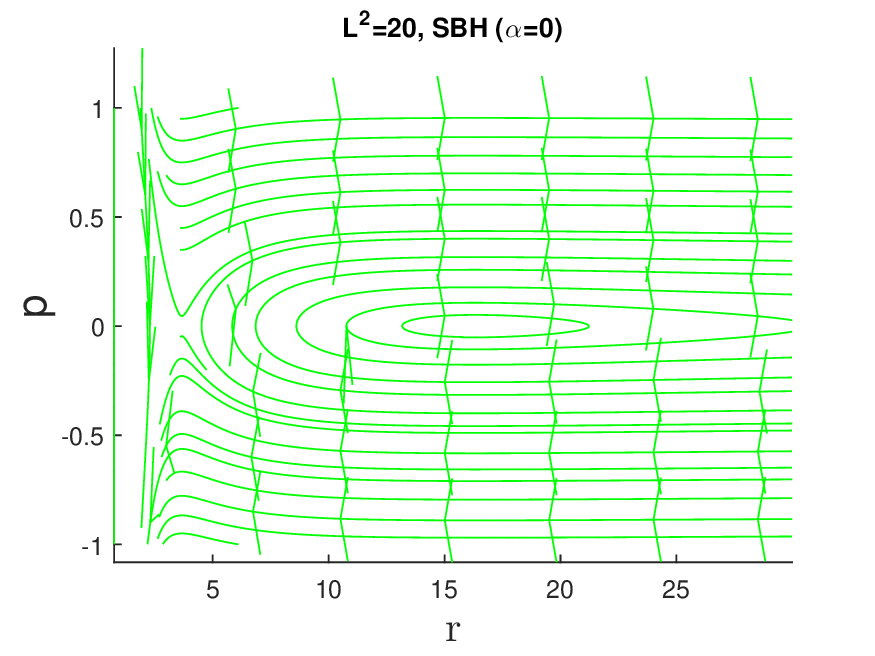}\label{ppt5}}
 	\subfigure[]{\includegraphics[scale=0.5]{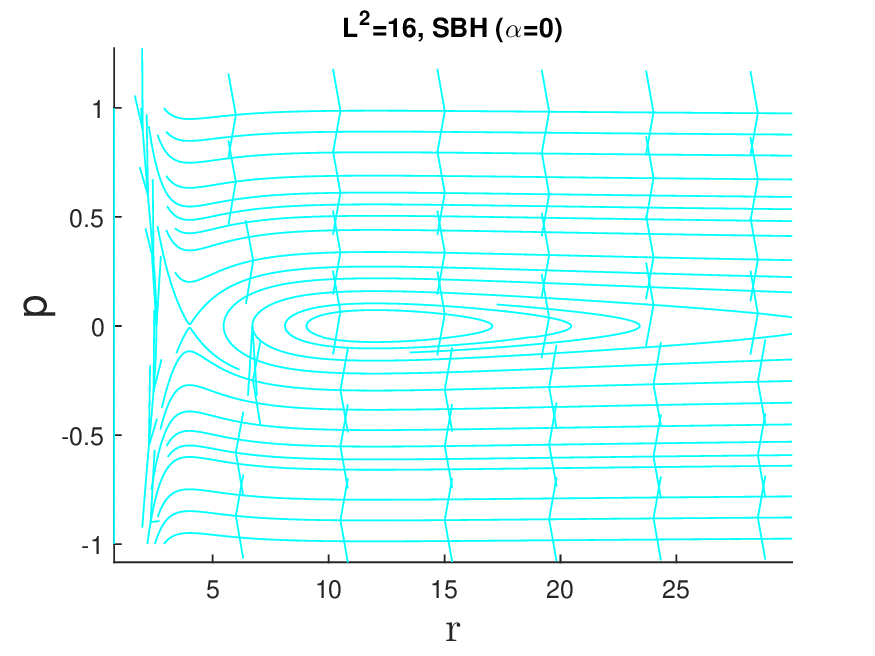}\label{ppt6}}
 	\caption {The phase portrait in $r-p$ plane for timelike geodesics by varying the angular momentum  $(L^{2}=16, 20)$ for different string cloud parameters and the pure SBH case. }\label{phaseporT}
 \end{figure*}

\begin{figure*}[htbp]
 	\centering
 	\subfigure[]{\includegraphics[scale=0.5]{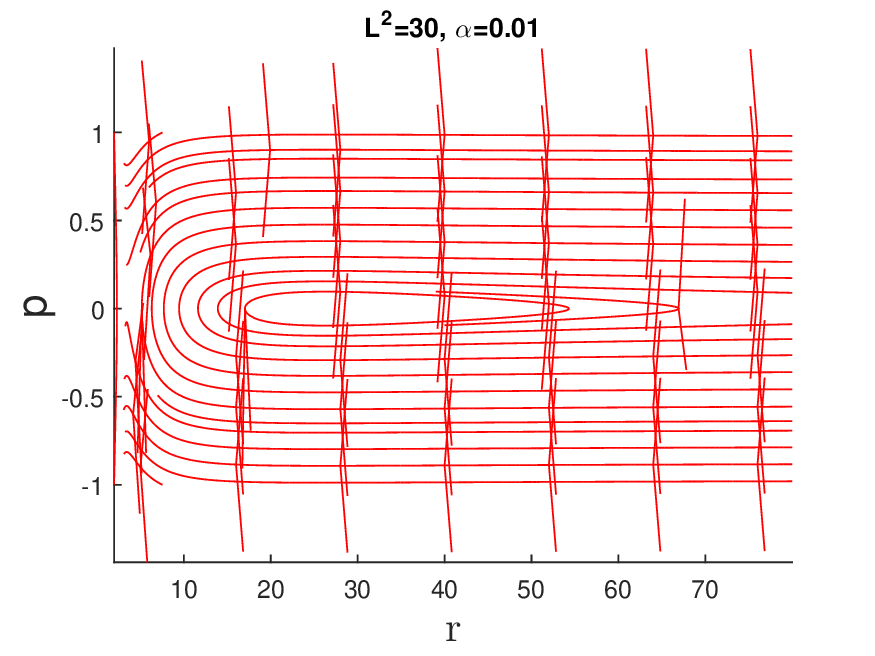}\label{ppt7}}
 	\subfigure[]{\includegraphics[scale=0.5]{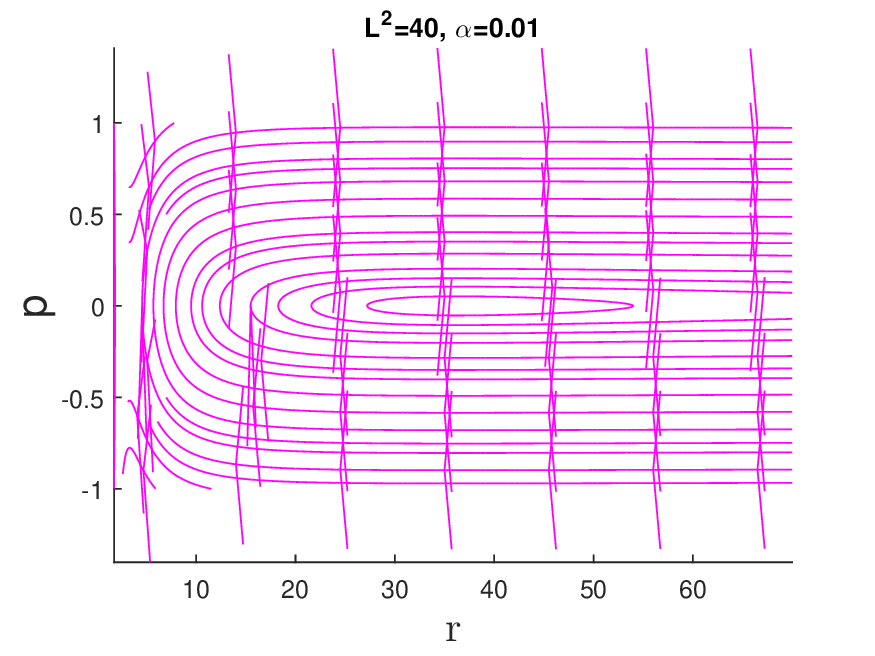}\label{ppt8}}
 	\subfigure[]{\includegraphics[scale=0.5]{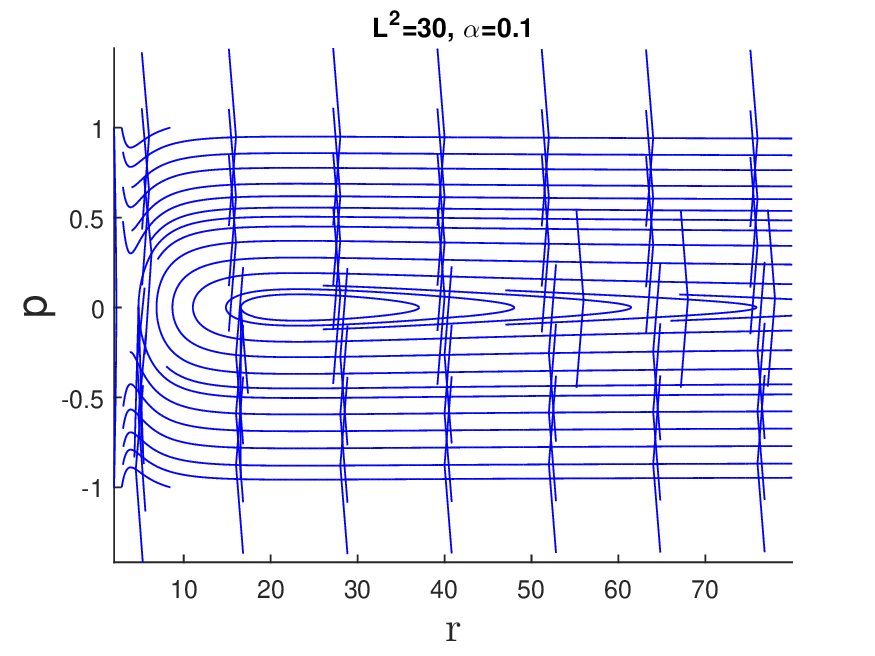}\label{ppt9}}
 	\subfigure[]{\includegraphics[scale=0.5]{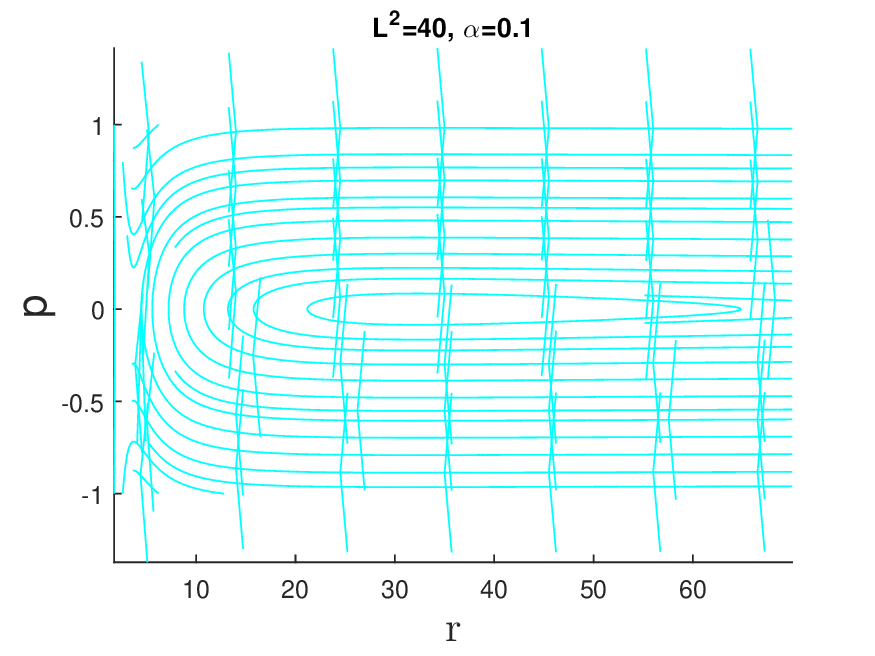}\label{ppt10}}
 	\subfigure[]{\includegraphics[scale=0.5]{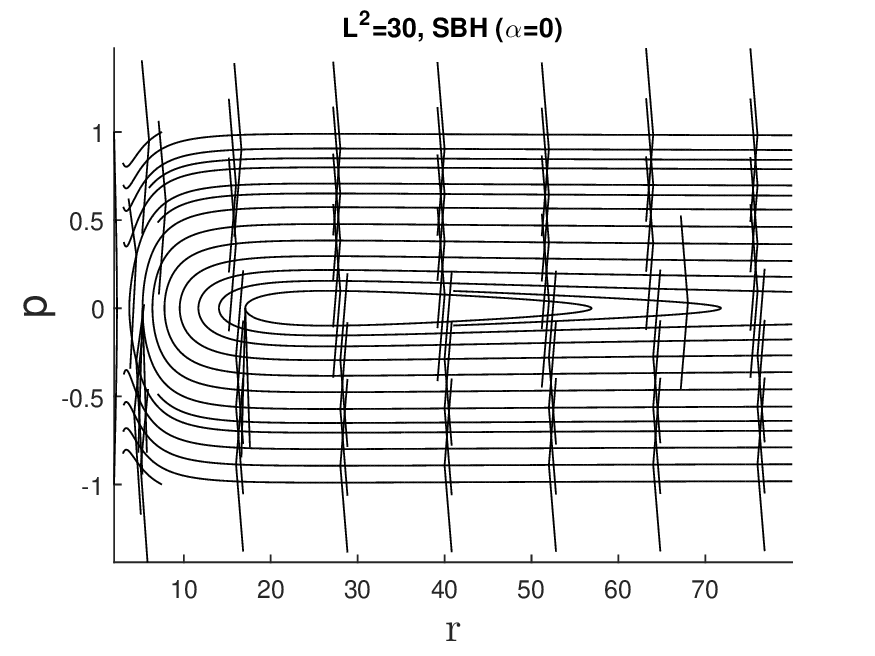}\label{ppt11}}
 	\subfigure[]{\includegraphics[scale=0.5]{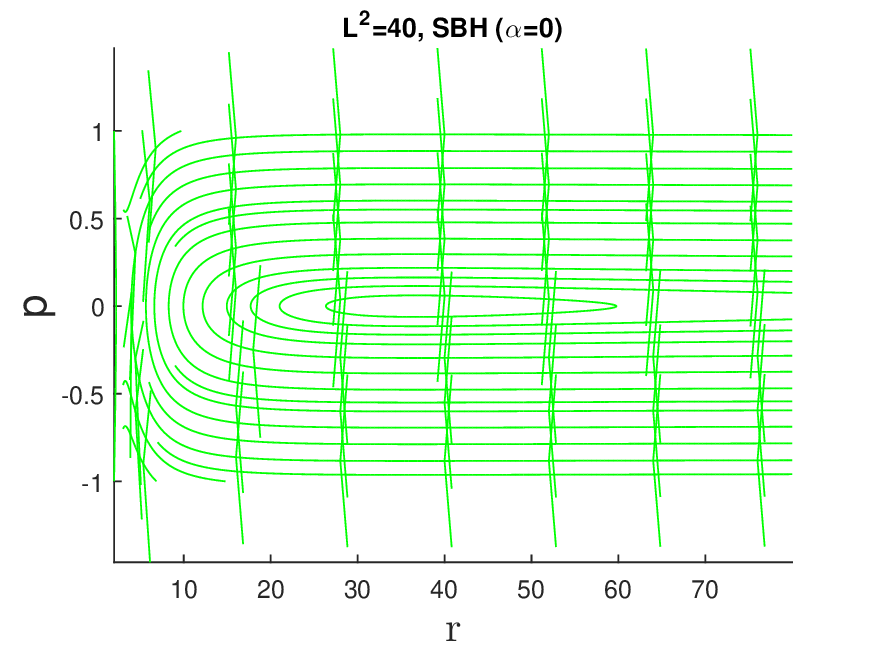}\label{ppt12}}
 	\caption {The phase portrait in $r-p$ plane for timelike geodesics by varying the angular momentum $(L^{2}=30, 40)$ for different string cloud parameters and the pure SBH case. }\label{phaseporT2}
\end{figure*}

\noindent The Lyapunov exponent at a fixed point for the timelike geodesics is thus given by  $\lambda_{T}=\pm \sqrt{-\left(V_{eff}^{T}(r_{0})\right)''}$, and the fixed points for each case are classified in \tablename{ \ref{table2}}. 
The phase portraits of these fixed points are visualized in \figurename{\ref{phaseporT}} and \figurename{\ref{phaseporT2}} and these fixed points are classified as either unstable saddle points or stable centres corresponding to the Lyapunov exponent eigenvalues, which are real and imaginary, respectively. Thus for different values of string cloud parameters and  angular momentum, it is observed that
\begin{itemize}
    \item At a fixed string cloud parameter, when we decrease the angular momentum, the position of the unstable circular orbit shifts to the higher value of $r$ while the position of the stable circular orbit shifts to the lower value of $r$.
    \item At a fixed angular momentum, if we increase the string cloud parameter, the unstable circular orbit shifts to a higher value of $r$ while the stable circular orbit shifts to a lower value of $r$.
    \item In case of SBH  ($\alpha=0$) for a particular value of angular momentum (e.g. $L^{2}=16$), the unstable circular orbit shifts to lower radius at the fixed point $(4,0)$ while the stable orbit shifts to a higher radius at the fixed point $(12,0)$.
\end{itemize}

\subsection{Stability of Null Geodesics}
We investigate the motion of light or photons around the SCSBH spacetime to observe the effect of the string cloud on the stability of null geodesics. The effective potential for null geodesics (taking $\delta=0$) is expressed as
\begin{equation}
V_{eff}^{N}(r)=  \frac{1}{2}\left(\frac{L^{2}}{r^{2}}\right) \left[1-\frac{2M}{r}-\alpha\right]  .\label{effN}
\end{equation}

\begin{figure*}[htbp]
	\centering
	\subfigure[]{\includegraphics[scale=0.9]{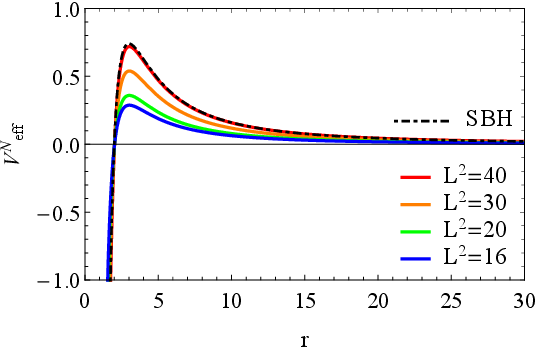}}
	\subfigure[]{\includegraphics[scale=0.9]{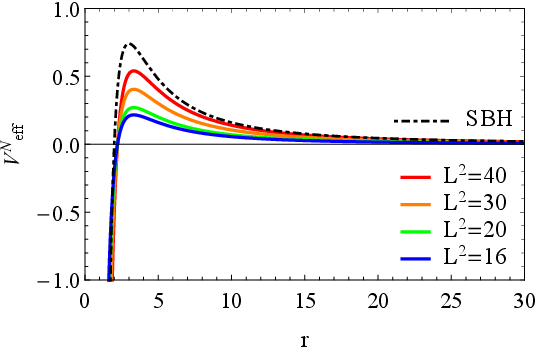}}
	\caption {Behaviour of the effective potential $V_{eff}^{N}$ versus $r$ by varying angular momentum and different string cloud parameters (a) $\alpha=0.01$, (b) $\alpha=0.1$, and for SBH considering $L^{2}=40$. Here we also fixed $M=1$. }\label{effectiveNull}
\end{figure*}

\noindent The effective potential $V_{eff}^{N}$ is plotted against the radius of orbits $(r)$ in \figurename{\ref{effectiveNull}}. It shows that the maxima of the potential decrease as we decrease the angular momentum, but \emph{increase} as we decrease the string cloud parameter, since substitution of the photon-sphere radius $r_{*}=3M/(1-\alpha)$ into Eq.~\eqref{effN} gives $V_{eff}^{N}(r_{*})=L^{2}(1-\alpha)^{3}/(54M^{2})$, which scales as $(1-\alpha)^{3}$ and therefore grows as $\alpha\to 0$.

Further, we obtain only one fixed point $(r_{*},0) $ for a particular value of string cloud parameter $\alpha$ i.e. (3.0, 0) for $\alpha=0.0$ (SBH), (3.0303, 0) for $\alpha=0.01$, (3.3333, 0) for $\alpha=0.1$ and (4, 0) for $\alpha=0.25$ (see \tablename{ \ref{table3}}). It implies that the shifting of the fixed point of the effective potential is independent of the angular momentum but depends on the parameter $\alpha$. Analytically, $r_{*}=3M/(1-\alpha)$, which reproduces the three tabulated values for $M=1$. \\
\noindent To check the Lyapunov stability of null circular geodesics, we find the second-order derivative at the fixed point $\left(V_{eff}^{N}(r_{*})\right)''$ and obtain the corresponding Lyapunov exponent ($\lambda_{N}$). The circular geodesics at each fixed point are found to be unstable saddle points, as the corresponding Lyapunov exponent is real.

\begin{table*}[tbhp]
	\caption{ Lyapunov stability of null geodesics at fixed points ($r_{*},0$) for different values of parameter
		$\alpha$.}
	\centering
	\begin{tabular}{|c|c|c|c|c|}
		\hline
		String cloud parameter & Fixed point ($r_{*},0$) & $\left(V_{eff}^{N}(r_{*})\right)''$ & $\lambda_{N}$  & 	Lyapunov stability  \\  [0.5ex]
		\hline
		$\alpha=0~(SBH)$&	(3, 0) & Negative & Real & Unstable saddle \\
		\hline
		$\alpha=0.01$&	(3.0303, 0) & Negative & Real & Unstable saddle \\
		\hline
		$\alpha=0.1$&	(3.3333, 0) & Negative & Real & Unstable saddle \\
		\hline
		$\alpha=0.25$&	(4, 0) & Negative & Real & Unstable saddle \\
		\hline
	\end{tabular}
	\label{table3}
\end{table*}

 \begin{figure*}[htbp]
 	\centering
 	\subfigure[]{\includegraphics[scale=0.5]{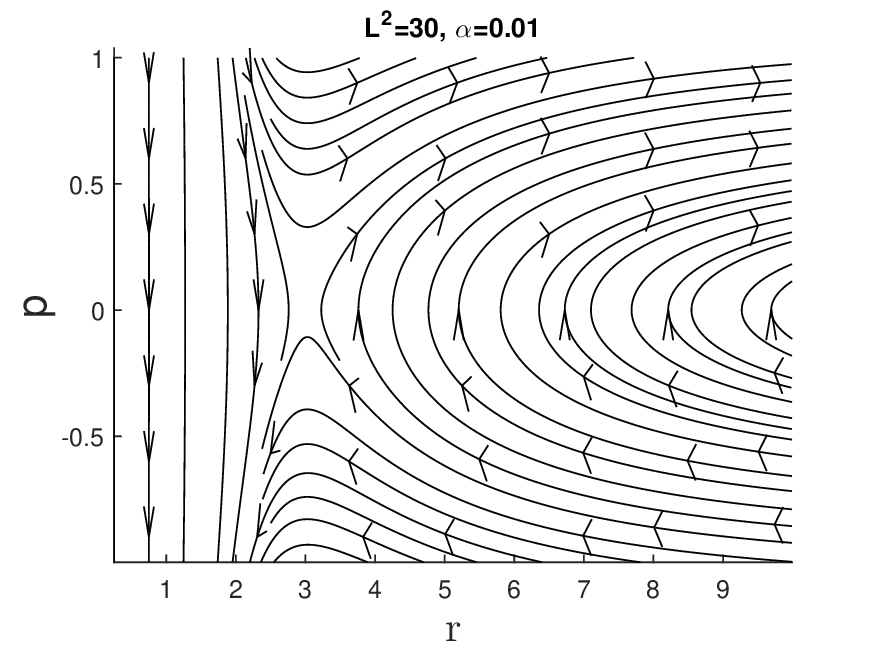}}
 	\subfigure[]{\includegraphics[scale=0.5]{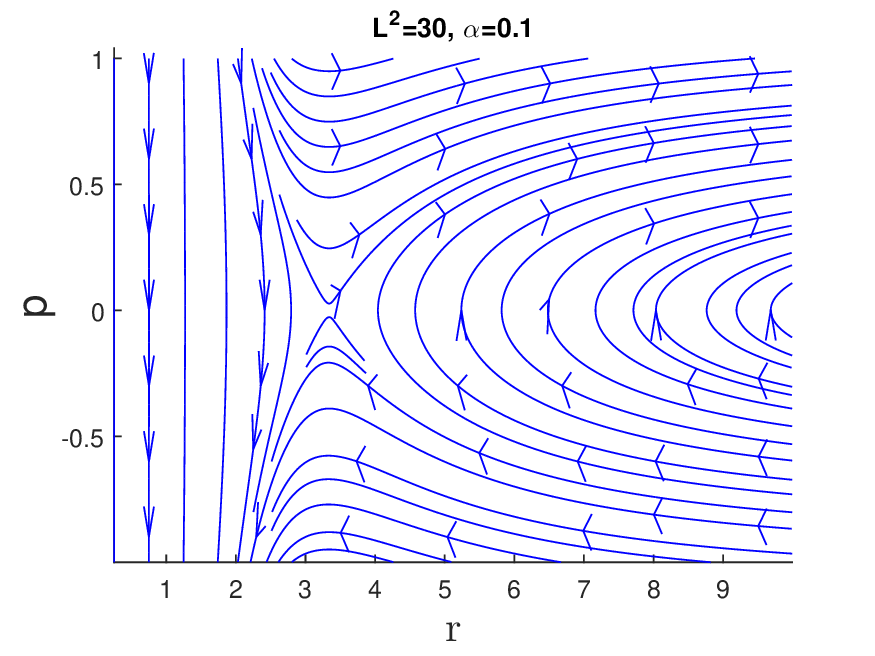}}
     \subfigure[]{\includegraphics[scale=0.5]{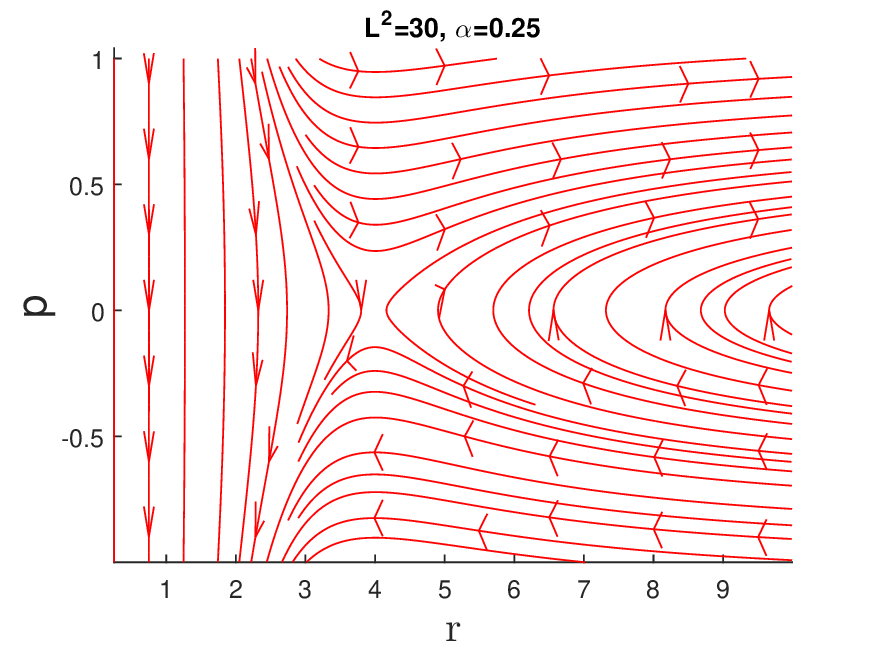}}
     \subfigure[]{\includegraphics[scale=0.5]{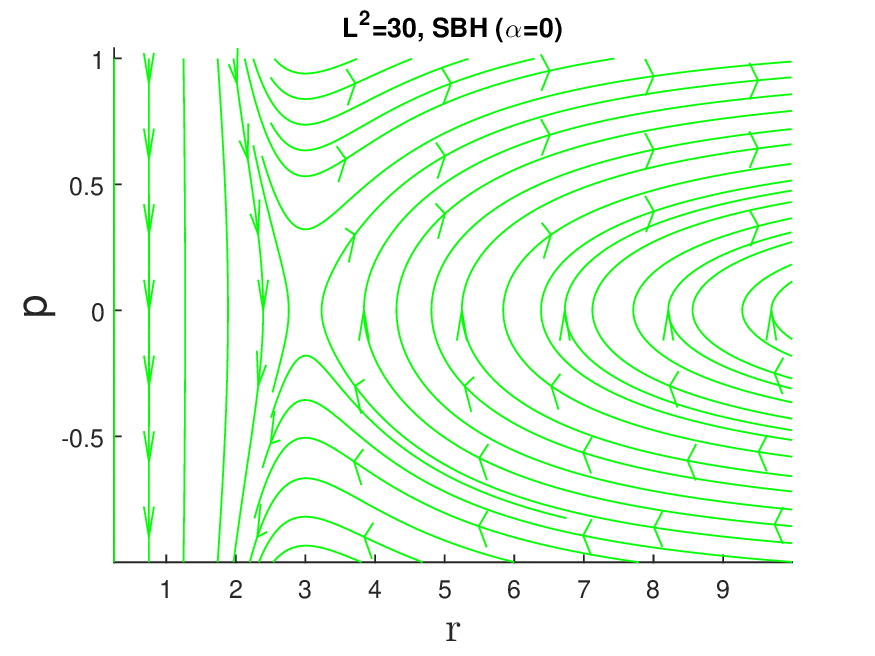}}
	\caption {Phase-portrait for null geodesics by varying angular momentum and different string cloud parameters $\alpha$ and SBH.} \label{ppnull}
 \end{figure*}

In order to decide whether the circular orbits at fixed points are stable or unstable, we have visualized the phase portrait in \figurename{\ref{ppnull}}. It is observed that the fixed point for each string cloud parameter $\alpha$ is an unstable saddle point. Consequently, the circular orbits of photons at the fixed point are found to be unstable i.e. no stable circular orbits exist for null geodesics in SCSBH spacetime.

\section{Summary and conclusions}
We have investigated the influence of the string cloud parameter on the stability of geodesics around SCSBH spacetime using the Lyapunov exponent and phase-portrait analysis of circular geodesics.
We analyse the dependence of the timelike effective potential $V_{eff}^{T}$ on the radial coordinate for various values of the angular momentum. It is observed that for a small string cloud parameter, the spacetime geometry remains nearly unchanged, and the effective potential of the SCSBH closely matches that of the SBH. In contrast, when $\alpha = 0.1$, the string cloud significantly modifies the spacetime geometry, lowering the potential barrier and enabling test particles to approach the black hole more closely than in the SBH spacetime.
\\
Further, we characterised the fixed point of $V_{eff}^{T}$ as a linearly stable centre or an unstable saddle by evaluating the Lyapunov exponent. Next, by depicting phase portraits, we observed that for a fixed string cloud parameter, decreasing the angular momentum shifts the unstable circular orbit to larger radii and the stable circular orbit to smaller radii. Similarly, for a fixed angular momentum, increasing the string cloud parameter moves the unstable circular orbit outward while shifting the stable circular orbit inward. \\
Furthermore, we also examine the effective potential $V_{eff}^{N}$ for the null case and find that the value at maximum scales as $(1-\alpha)^{3}$ and therefore decreases with increasing $\alpha$. In contrast, the maxima decrease with decreasing angular momentum. It is further observed that the fixed points associated with null geodesics are unstable saddle points, as indicated by the real-valued Lyapunov exponent. Consequently, the corresponding circular null geodesics are unstable, which is consistent with the phase portraits exhibiting only saddle-type behavior around these fixed points.
A comparison with the limiting case of a vanishing string cloud parameter (i.e., SBH case) has also been carried out to elucidate the role of the string cloud in modifying the geodesic dynamics.

\section*{Acknowledgements}
HN acknowledges the financial support provided by the Science and Engineering Research Board (SERB), New	Delhi, through grant number CRG/2023/-008980.
\bibliographystyle{unsrt}
\bibliography{SCSBHARxiv}

@Article{Giri:2022fjy,
  author        = {Giri, Shobhit and Nandan, Hemwati and Joshi, Lokesh Kumar and Maharaj, Sunil D.},
  journal       = {Eur. Phys. J. C},
  title         = {{Stability analysis of circular orbits around a traversable wormhole with massless conformally coupled scalar field}},
  year          = {2022},
  number        = {4},
  pages         = {298},
  volume        = {82},
  archiveprefix = {arXiv},
  doi           = {10.1140/epjc/s10052-022-10274-4},
  eprint        = {2204.10988},
  primaryclass  = {gr-qc},
}

@article{Abolghasem2013STABILITYOC,
  title={STABILITY OF CIRCULAR ORBITS IN SCHWARZSCHILD SPACETIME},
  author={Hossein Abolghasem},
  journal={International journal of differential equations and applications},
  year={2013},
  volume={12},
  url={https://api.semanticscholar.org/CorpusID:120214865}
}

@book{Becker:2006dvp,
    author = "Becker, K. and Becker, M. and Schwarz, J. H.",
    title = "{String theory and M-theory: A modern introduction}",
    doi = "10.1017/CBO9780511816086",
    isbn = "978-0-511-25486-4, 978-0-521-86069-7, 978-0-511-81608-6",
    publisher = "Cambridge University Press",
    month = "12",
    year = "2006"
}

@article{HenryTye:2006uv,
    author = "Henry Tye, S. -H.",
    title = "{Brane inflation: String theory viewed from the cosmos}",
    eprint = "hep-th/0610221",
    archivePrefix = "arXiv",
    journal = "Lect. Notes Phys.",
    volume = "737",
    pages = "949--974",
    year = "2008"
}

@article{Hindmarsh:1994re,
    author = "Hindmarsh, M. B. and Kibble, T. W. B.",
    title = "{Cosmic strings}",
    eprint = "hep-ph/9411342",
    archivePrefix = "arXiv",
    reportNumber = "SUSX-TP-94-74, IMPERIAL-TP-94-95-5, NI-94025",
    doi = "10.1088/0034-4885/58/5/001",
    journal = "Rept. Prog. Phys.",
    volume = "58",
    pages = "477--562",
    year = "1995"
}

@article{Letelier:1979ej,
    author = "Letelier, P. S.",
    title = "{CLOUDS OF STRINGS IN GENERAL RELATIVITY}",
    doi = "10.1103/PhysRevD.20.1294",
    journal = "Phys. Rev. D",
    volume = "20",
    pages = "1294--1302",
    year = "1979"
}

@article{Richarte:2007bx,
    author = "Richarte, Martin G. and Simeone, Claudio",
    title = "{Traversable wormholes in a string cloud}",
    eprint = "0711.2297",
    archivePrefix = "arXiv",
    primaryClass = "gr-qc",
    doi = "10.1142/S0218271808012759",
    journal = "Int. J. Mod. Phys. D",
    volume = "17",
    pages = "1179--1196",
    year = "2008"
}

@article{Kibble:1976sj,
    author = "Kibble, T. W. B.",
    title = "{Topology of Cosmic Domains and Strings}",
    reportNumber = "ICTP/75/5",
    doi = "10.1088/0305-4470/9/8/029",
    journal = "J. Phys. A",
    volume = "9",
    pages = "1387--1398",
    year = "1976"
}

@article{Vilenkin:1981iu,
    author = "Vilenkin, A.",
    title = "{Cosmological Density Fluctuations Produced by Vacuum Strings}",
    doi = "10.1103/PhysRevLett.46.1496",
    journal = "Phys. Rev. Lett.",
    volume = "46",
    pages = "1169--1172",
    year = "1981",
    note = "[Erratum: Phys.Rev.Lett. 46, 1496 (1981)]"
}

@article{Mitchell:1987th,
    author = "Mitchell, David and Turok, Neil",
    title = "{Statistical Properties of Cosmic Strings}",
    reportNumber = "IMPERIAL-TP-87-88-19",
    doi = "10.1016/0550-3213(87)90626-2",
    journal = "Nucl. Phys. B",
    volume = "294",
    pages = "1138--1163",
    year = "1987"
}

@article{Larsen:1997ge,
    author = "Larsen, Finn",
    title = "{A String model of black hole microstates}",
    eprint = "hep-th/9702153",
    archivePrefix = "arXiv",
    reportNumber = "UPR-0735-T, UPR-735-T",
    doi = "10.1103/PhysRevD.56.1005",
    journal = "Phys. Rev. D",
    volume = "56",
    pages = "1005--1008",
    year = "1997"
}

@article{Mazharimousavi:2009tp,
    author = "Mazharimousavi, S. Habib and Gurtug, O. and Halilsoy, M.",
    title = "{Theorem to generate Einstein-Non Linear Maxwell Fields}",
    eprint = "0911.1919",
    archivePrefix = "arXiv",
    primaryClass = "gr-qc",
    doi = "10.1088/0264-9381/27/20/205022",
    journal = "Class. Quant. Grav.",
    volume = "27",
    pages = "205022",
    year = "2010"
}

@article{Batool:2017bjw,
    author = "Batool, Mahwish and Hussain, Ibrar",
    editor = "De Paolis, Francesco and Gurzadyan, Vahe G. and Siddiqui, Azad A. and Nucita, Achille",
    title = "{Null and timelike geodesics of the Schwarzschild black hole with string cloud background}",
    doi = "10.1142/S021827181741005X",
    journal = "Int. J. Mod. Phys. D",
    volume = "26",
    number = "5",
    pages = "1741005",
    year = "2017"
}

@article{Herscovich:2010vr,
    author = "Herscovich, Estanislao and Richarte, Martin G.",
    title = "{Black holes in Einstein-Gauss-Bonnet gravity with a string cloud background}",
    eprint = "1004.3754",
    archivePrefix = "arXiv",
    primaryClass = "hep-th",
    doi = "10.1016/j.physletb.2010.04.065",
    journal = "Phys. Lett. B",
    volume = "689",
    pages = "192--200",
    year = "2010"
}

@article{MoraisGraca:2017nlv,
    author = "Morais Gra\c{c}a, J. P. and Lobo, Iarley P. and Salako, I. G.",
    title = "{Cloud of strings in $f(R)$ gravity}",
    eprint = "1708.08398",
    archivePrefix = "arXiv",
    primaryClass = "gr-qc",
    doi = "10.1088/1674-1137/42/6/063105",
    journal = "Chin. Phys. C",
    volume = "42",
    number = "6",
    pages = "063105",
    year = "2018"
}

@article{Lee:2014dha,
    author = "Lee, Tae-Hun and Baboolal, Dharmanand and Ghosh, Sushant G.",
    title = "{Lovelock black holes in a string cloud background}",
    eprint = "1409.2615",
    archivePrefix = "arXiv",
    primaryClass = "gr-qc",
    doi = "10.1140/epjc/s10052-015-3515-5",
    journal = "Eur. Phys. J. C",
    volume = "75",
    number = "7",
    pages = "297",
    year = "2015"
}

@article{Ghosh:2014dqa,
    author = "Ghosh, Sushant G. and Maharaj, Sunil D.",
    title = "{Cloud of strings for radiating black holes in Lovelock gravity}",
    eprint = "1409.7874",
    archivePrefix = "arXiv",
    primaryClass = "gr-qc",
    doi = "10.1103/PhysRevD.89.084027",
    journal = "Phys. Rev. D",
    volume = "89",
    number = "8",
    pages = "084027",
    year = "2014"
}

@article{Ganguly:2014cqa,
    author = "Ganguly, Apratim and Ghosh, Sushant G. and Maharaj, Sunil D.",
    title = "{Accretion onto a black hole in a string cloud background}",
    eprint = "1409.7872",
    archivePrefix = "arXiv",
    primaryClass = "gr-qc",
    doi = "10.1103/PhysRevD.90.064037",
    journal = "Phys. Rev. D",
    volume = "90",
    number = "6",
    pages = "064037",
    year = "2014"
}

@article{Sharif:2015oua,
    author = "Sharif, M. and Iftikhar, Sehrish",
    title = "{Null Geodesics and Strong Field Gravitational Lensing in a String Cloud Background}",
    doi = "10.1155/2015/635625",
    journal = "Adv. High Energy Phys.",
    volume = "2015",
    pages = "635625",
    year = "2015",
    note = "[Erratum: Adv.High Energy Phys. 2015, 219762 (2015)]"
}

@article{boehmer2012jacobi,
  title={Jacobi stability analysis of dynamical systems—applications in gravitation and cosmology},
  author={Boehmer, CG and Harko, T and Sabau, SV},
  journal={Advances in Theoretical and Mathematical Physics},
  volume={16},
  number={4},
  pages={1145--1196},
  year={2012},
  publisher={International Press of Boston}
}

@article{abolghasem2012jacobi,
  title={Jacobi stability of circular orbits in a central force},
  author={Abolghasem, Hossein},
  journal={Journal of Dynamical Systems and Geometric Theories},
  volume={10},
  number={2},
  pages={197--214},
  year={2012},
  publisher={Taylor \& Francis}
}

@incollection{cartan2016observations,
  title={Observations sur le m{\'e}moire pr{\'e}c{\'e}dent},
  author={Cartan, {\'E}lie},
  booktitle={DD Kosambi},
  pages={71--74},
  year={2016},
  publisher={Springer}
}

@incollection{kosambi2016parallelism,
  title={Parallelism and path-spaces},
  author={Kosambi, Damodar Dharmananda},
  booktitle={DD Kosambi},
  pages={59--70},
  year={2016},
  publisher={Springer}
}

@article{PhysRevLett.55.1082,
  title = {Measurement of the Lyapunov Spectrum from a Chaotic Time Series},
  author = {Sano, M. and Sawada, Y.},
  journal = {Phys. Rev. Lett.},
  volume = {55},
  issue = {10},
  pages = {1082--1085},
  numpages = {0},
  year = {1985},
  month = {Sep},
  publisher = {American Physical Society},
  doi = {10.1103/PhysRevLett.55.1082},
  url = {https://link.aps.org/doi/10.1103/PhysRevLett.55.1082}
}

@article{Cardoso:2008bp,
    author = "Cardoso, Vitor and Miranda, Alex S. and Berti, Emanuele and Witek, Helvi and Zanchin, Vilson T.",
    title = "{Geodesic stability, Lyapunov exponents and quasinormal modes}",
    eprint = "0812.1806",
    archivePrefix = "arXiv",
    primaryClass = "hep-th",
    doi = "10.1103/PhysRevD.79.064016",
    journal = "Phys. Rev. D",
    volume = "79",
    number = "6",
    pages = "064016",
    year = "2009"
}

@Article{AMLyapunov,
  author    = {A. M. LYAPUNOV},
  journal   = {International Journal of Control},
  title     = {The general problem of the stability of motion},
  year      = {1992},
  number    = {3},
  pages     = {531--534},
  volume    = {55},
  doi       = {10.1080/00207179208934253},
  eprint    = {https://doi.org/10.1080/00207179208934253},
  publisher = {Taylor \& Francis},
  url       = {https://doi.org/10.1080/00207179208934253},
}

@article{balan2009berwald,
  title={Berwald-Moor metrics and structural stability of conformally-deformed geodesic SODE.},
  author={Balan, Vladimir and Nicola, Ileana Rodica},
  journal={Applied Sciences},
  volume={11},
  year={2009}
}

@article{boehmer2010nonlinear,
  title="{Nonlinear stability analysis of the Emden--Fowler equation}",
  author={Boehmer, Christian G and Harko, Tiberiu},
  journal={Journal of Nonlinear Mathematical Physics},
  volume={17},
  number={4},
  pages={503--516},
  year={2010},
  publisher={Taylor \& Francis}
}

@article{sabau2005some,
  title={Some remarks on Jacobi stability},
  author={Sabau, SV},
  journal={Nonlinear Analysis: Theory, Methods \& Applications},
  volume={63},
  number={5-7},
  pages={e143--e153},
  year={2005},
  publisher={Elsevier}
}

@article{harko2008jacobi,
  title={Jacobi stability of the vacuum in the static spherically symmetric brane world models},
  author={Harko, T and Sabau, VS},
  journal={Physical Review D},
  volume={77},
  number={10},
  pages={104009},
  year={2008},
  publisher={APS}
}

\end{document}